\documentclass[twocolumn, aps, pra, superscriptaddress]{revtex4-1}
\pdfoutput=1
\usepackage[utf8]{inputenc}
\usepackage[english]{babel}
\usepackage[T1]{fontenc}
\usepackage{hyperref}
\usepackage{xurl}
\usepackage{graphicx}
\usepackage{amsfonts, amsmath, amsthm, amssymb} 
\usepackage{braket}
\usepackage{float}
\usepackage{amsthm}
\usepackage[all]{nowidow}
\usepackage{listings}
\usepackage{xcolor}

\begin{document}

\title{Quandoom - DOOM as a quantum circuit}

\author{Luke Mortimer}
\affiliation{ICFO-Institut de Ciencies Fotoniques, The Barcelona Institute of Science and Technology, 08860 Castelldefels, Spain}

\date{\today}

\begin{abstract}
\noindent
Since the early 2000s there has existed the meme that ``DOOM can run on anything''. Whether it be an ATM or a calculator, someone at some point has recompiled DOOM to run on it. Now the quantum computer finally joins the list. More specifically, this project represents the first level of DOOM loosely rewritten using Hadamards and Toffolis which, despite being a universal gate set, has been designed in such a way that it's classically simulable, able to reach 10-20 frames per second on a laptop. The circuit uses 72,376 total qubits and at least 80 million gates, thus it may have use as a benchmark for quantum simulation software.
\end{abstract}

\maketitle

\section{Introduction}

Whilst not the first 3D video game, DOOM was incredibly influential at the time of its release in 1993. Often called ``the father of first person shooters'' \cite{doom}, it led to a number of copycats \cite{doom2} and its design choices have heavily influenced the video game industry since. Owing to this importance, as well as its low computational requirements and the fact that the engine code was made open-source \cite{id2024id}, it quickly became an internet meme that ``DOOM can run on anything''. Some extreme examples include a pregnancy test \cite{pregnancy} or human gut bacteria \cite{ecoli}, whilst many more examples can be found on the Reddit community ``/r/itrunsdoom'' \cite{subreddit}.

Meanwhile, quantum computing still remains in the so-called NISQ era \cite{preskill2018quantum}, where we are generally limited to low-depth circuits with a small number of qubits. At the time of writing there has yet to be any demonstrated practical advantage of quantum computing. In fact, it remains dubious whether quantum computing will ever reach a point whereby it outperforms classical computers in anything other than hyper-specific test problems without practical use \cite{me}.

For classical computers, there exist a number of standards for representing the list of operations we want to perform. The most widely used of these is the assembly language which, although the exact list of instructions may vary between computational architectures, defines a list of operations that can be applied to classical bit registers. For instance, one might perform ``MOV AL, 61h'' to move the hexidecimal value 61h into register AL. Similarly we can define the quantum equivalent, with a common choice being OpenQASM, a Quantum ASseMbly language developed by IBM \cite{cross2022openqasm}. Such QASM files contain definitions of quantum registers and operations to be applied to said registers. For instance, one might perform ``X q[0]'' to apply an X gate on the the first qubit in the register ``q''.

There have already been some examples of ``quantum video games'', for instance by Dr. James Wootton \cite{games}. I also previous developed a quantum video game as a QASM file: a dungeon crawler known as ``The Quantum Tunnels'' \cite{tunnels}. However, in general there are very few examples due to the fact that the there is no advantage to be gained. But regardless, in this work we add one more to the list of quantum games with the release of Quandoom - a quantum port of the first level of DOOM.

The basic game loop of Quandoom is as follows: we take our quantum state, set the input qubits based on the keys the user is pressing, then apply the quantum circuit to the state using a custom-built simulator, then measure the last 64000 qubits in the Z basis and render them as a 320x200 grid of binary pixels. The process then repeats, keeping the same state, only measuring/resetting the screen and input qubits. The circuit itself \cite{quandoom} and the C++ code used to generate it \cite{engine} are both open source. If a C++ source file (i.e. ``something.cpp'') is referenced in this document, it refers to the latter repository.

\begin{figure}
    \centering
    \includegraphics[width=0.99\linewidth]{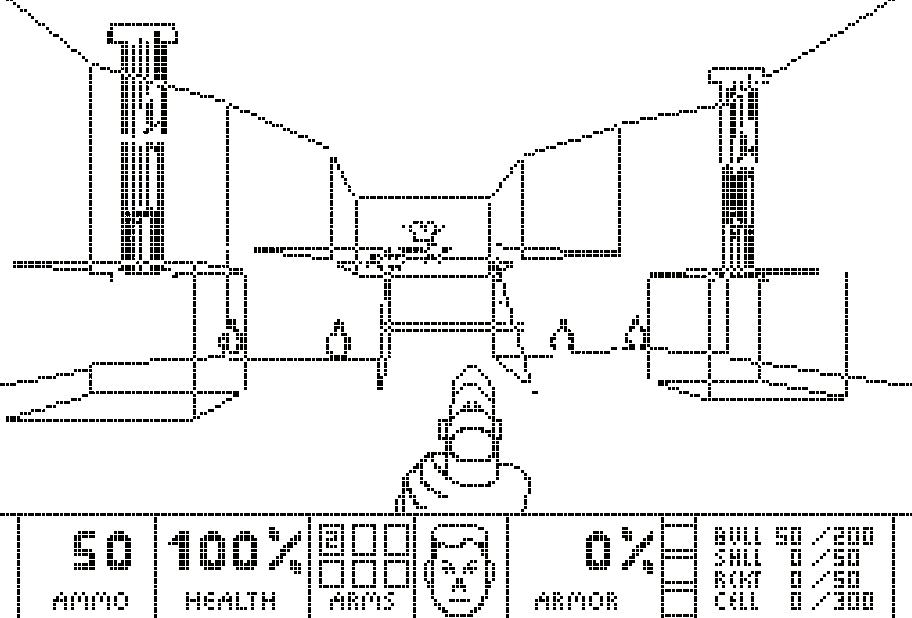}
    \caption{A screenshot of Quandoom being simulated. Each pixel is a binary value representing the measurement of one of the output qubits. No classical logic is performed, all rendering and game logic is done in the quantum framework using Toffoli and Hadamard gates. The color has been inverted to better suit this document.}
    \label{fig:screenshot}
\end{figure}

\section{Registers and Basic Operations}

We begin with list of registers - the list of qubits which we will maintain between runs. Each register represents a quantity, for instance we have a register called ``health'' which stores the health of the player. Such quantities are all stored as an integer in two's complement form, thus allowing negatives. I should really have done it with floating points instead, but at some point I was in too deep with the integer thing and could not be bothered to rewrite the entire code. A small subset of the registers defined in registers.cpp is given as Table \ref{tab:registers}

\begin{table}
    \centering
    \begin{tabular}{c|c|c}
         Register & Qubits & Usage \\
         \hline
         health & 9 & the player's health count \\
         ammoBullets & 8 &  the player's ammo count \\
         enemyDrawAction & 4 & used for enemy animations \\
         weaponFrame & 3 & used for player weapon animations \\
         lineXWorldStart & 16 & line start x in world space \\
         lineXWorldEnd & 16 & line end x in world space \\
         lineXScreenStart & 10 & line start x in screen space \\
         lineXScreenEnd & 10 & line end x in screen space \\
         pixelToDrawX & 10 & x location of the pixel to draw \\
         cosTheta & 23 & cached 32 * cosine of player angle \\
         outputPixels & 64000 & screen qubits to render \\
         ancilla & 6986 & extra qubits used in calculations \\
    \end{tabular}
    \caption{A small subset of the qubit registers used in Quandoom, including their qubit size and a small (i.e. fitting in the column) description of their usage.}
    \label{tab:registers}
\end{table}

To act on these registers we define a number of basic operations - addition, subtract, division, multiplication, comparison, checking equality. It is from these that we will build up all of the higher-level functions. These subroutines will all be implemented using the most general form of the X gate, the multiple-controlled multiple-target X gate, which flips a number of qubits only if all of its control qubits are 1. It was pointed out after the initial release of the QASM file that this general form of the gate is not officially in the OpenQASM standard, however I find that ridiculous and chose to use it regardless. Thus although technically the file is not true QASM, owing as well as to the fact that some register lists are abbreviated for the sake of file size, the circuit remains a valid quantum circuit despite IBM's arbitrary standards.

Part of the reason that the quantum circuit is so inefficient compared to the classical algorithm is due to the reversibility requirement of quantum circuits. Whilst classically we can do something like ``add 1 to a number, if it goes above 5 then set it back to 5'', this would be an irreversible operation since then one could not determine whether the previous number was 4 or 5 (since they both map to 5). As such, the quantum version of the above algorithm has to be ``add 1 to a number, wrapping back around to the minimum if we go above 5''. As one might imagine, this creates a number of challenges, although in many cases this can be avoided by placing the result in a new register rather than overwriting the input.

We will not go into the specific implementations of all of the various basic operations here, to view the specifics please see binary.cpp, but basically we now have a number of C++ subroutines which take some of our registers as arguments and then add the corresponding QASM to our final circuit. For instance, we have \textit{flipIfLessOrEqualTC(regToFlip, regToCheck, regToCheck2)} which flips a register if the value of the first check register is less than that of the second check register, done by subtracting the registers and then checking the sign. These basic routines are far from optimal, but eh, I wasn't aiming for optimality, the aim of the project was just to get something that works. But if you want to try to improve the performance or even just to mess around with the code, I encourage you to fork the GitHub repository and have a go.

\section{Rendering}

Given a wide possible range of operations acting on our registers, we begin to implement the subroutines allowing for the drawing and rendering to the screen qubits, all implemented in draw.cpp. The lowest operation in the render stack is the subroutine ``drawPixel'' which flips a pixel based on the values of the registers ``pixelToDrawX'' and ``pixelToDrawY''. This is a simple one-to-one map, such that if the X register is 286 and the Y register is 124, we should flip qubit 39966 of the screen register ($320\times 124 + 286$). Due to the reversible nature, we can only flip the output qubits, never set them. This creates a wireframe x-ray visual, such that overlapping lines create a negative where they overlap, and objects can be seen behind other objects. It would not be impossible to fix these issues, one would need to pre-render each object (keeping them all in memory) and then only actually draw to screen the ones which are in front and don't overlap, before uncomputing each of the pre-renders. Or alternatively there may be a solution using a number of screens registers acting as different Z-buffers, which can then be merged into one final screen.

With the ability to draw a pixel, we now move on to drawing a line. To do this, we create some temporary ancilla registers (which we will release once we have uncomputed everything in them, to be used again later). We create a new X and Y register, onto which we copy the start location of the line. We then add values to these until we reach the end point of the line, at each step swapping the register with the pixelToDraw registers and calling the drawPixel gate before swapping back. Whilst in this case we do have a loop, it is a loop of fixed size in which we simply draw each pixel conditioned on a register stating whether we have reached the end of the line or not. It is not possible to do a loop of custom length in a quantum circuit unless one takes measurements in the middle of the circuit (which we do not).

A sample of code from the drawLine subroutine is given as follows, in which we also demonstrate the ability to manage recordings (i.e. record a section of code, then add the inverse when uncompute is called) as well as the use of left shifts to keep everything as integers (i.e. that 0.5 is instead represented as 16, knowing that we will divide again by 32 later). Here \textit{maxDiag} is a fixed value representing the maximum line size ($=\sqrt{320^2+200^2}$).

\definecolor{backcolour}{rgb}{0.95,0.95,0.95}
\lstdefinestyle{mystyle}{
    backgroundcolor=\color{backcolour},   
    basicstyle=\ttfamily\footnotesize,
    breakatwhitespace=false,         
    breaklines=true,                 
    captionpos=b,                    
    keepspaces=true,                 
    numbers=left,                    
    numbersep=5pt,                  
    showspaces=false,                
    showstringspaces=false,
    showtabs=false,                  
    tabsize=2
}
\lstset{style=mystyle}

\begin{minipage}{\linewidth}
\begin{lstlisting}[language=C++]
rec = startRecording();
shiftLeft(subX, shiftBits);
divide(subX, subY, gradient, remainder); 
multiply(gradient, maxDiag, gradientTimesMax); 
copy(startX, movingX);
copy(startY, movingY);
shiftLeft(movingX, shiftBits); 
shiftLeft(movingY, shiftBits);
stopRecording(rec);
x(keepDrawing);
for (int i=0; i<maxDiag; i++) {
    add(movingX, gradient);
    add(movingY, shiftAmount);
    flipIfEqual(keepDrawing, ySubreg, endY);
    drawPixelIf(drawPixelName, keepDrawing);
}
subtract(movingY, shiftAmount*maxDiag);
subtract(movingX, gradientTimesMax, false);
uncomputeRecording(rec);
\end{lstlisting}
\end{minipage}

Now that we can draw a line in 2D (screen) space, we move to the much harder problem of trying to render a line in 3D (world) space. To do this we need to project a coordinate from a 3D space onto a 2D plane, in our case using a perspective projection algorithm. To do this, I simply went on the Wikipedia for 3D projection \cite{projection} and implemented the equations pretty much directly. There is no issue with reversibility here, since we keep the original point in memory and thus after applying to the 3D -> 2D map we can simply do it in reverse to uncompute everything, there is no need for it to be one-to-one since we keep the original in memory.

The most surprising thing here was the amount of ancilla qubits required for the projection subroutines - in projectCoord we use 16 registers each of 23 qubits = 368 qubits. These are used to cache the various multiplications/divisions that are required to implement the above equations. In renderLine we use even more - 64 registers of 23 qubits = 1472 qubits. This is due to needing to project two coordinates, then needing a large number of checks for frustum clipping. Whilst one could reduce this count a lot, by uncomputing all ancillas after every operation, this would result in an explosion in the gate count in exchange for using a few hundred less qubits. In our case, qubit count doesn't matter since we aren't going to simulate full qubits anyway, but several million extra gates could make the difference between playable in real time or not.

\section{Baking out of Laziness}

The concept of ``baking'' is a common idea in game development, where rather than calculating something on-the-fly we choose to precalculate things like static lighting or a certain physics collision in order to take pressure off the CPU when the user is actually playing the game. Quandoom also uses such techniques, but mostly out of laziness of rather than performance or playability.

The first example of this is sprite rendering. Normally in rendering one would project the top and bottom of a sprite to get the screen size of the sprite, then scale the sprite to the desired size before drawing it. The issue in the quantum case is that implementing a nearest neighbor scaling algorithm (or anything more advanced) would be quite complicated, and at the point of needing to implement it I was getting reasonably tired of the project (about 6 months of weekends spent on it) and thus it was decided to simply bake in every possible scaling of every sprite (every sprite each scaled to a height of n pixels, $1\le \forall n \le 200$). So, after we determine what size a given sprite should be, we simply run that corresponding quantum subroutine to draw that pre-scaled version of the sprite. This comes at the cost of the quantum circuit being a lot bigger and should be the first focus of anyone wanting to reduce the file size of the QASM file.

The second example is that of ray casting. Normally, in order to determine if an enemy can see a player, one could send a ``ray'' (i.e. traveling point) from the enemy towards the player and check if at any point it intersects with a blocking wall. In Quandoom, for similar reasons as the above paragraph, we simply separate every section of each room into a course grid (10x10 per room) and then precalculate whether each section of the grid can see each other section. This takes some time when generating the circuit, but in terms of gates it's reasonably efficient as it is only $(10x10)^2$ controlled X gates checking specific grid locations. That may sound like a lot, but compared to the rendering subroutines which each use several million, it's basically free.

\section{Level Design}

The level is split into several rooms, such that only one room is drawn at a time, based on the X and Y location of the player. This is a heavily simplified version of the binary-tree-based rendering of the original DOOM engine. There are also no secret rooms in Quandoom because it would have added gate count and file size and it was unlikely that many people would even play it, yet alone also know DOOM well enough to check the secrets.

The various rooms are defined in the file levels.cpp and were written by-hand based on a map of the first level of DOOM from the DOOM wiki \cite{wiki}. That is, the start and location of each wall was placed by me vaguely guessing the location and then generating the circuit and checking how it looks (i.e. tedious). Whilst I did consider trying to extract the level information from the DOOM ``.wad'' level file, due to the style of Quandoom's rendering quite a lot of walls were changed to make the rooms more convex, as well as many details removed for the sake of trying to keep the rendering smooth, thus manual tweaking was needed.

Enemies and pickups have their original locations, but health/armor/damage scaling was adjusted slightly to make it easier to play with the limited framerate and simplified enemy targeting mechanics. It was debated at first where the project should stop, should there be different difficulty levels? Settings and a main menu? An automap, secrets and an end-of-level screen? Music? The entire first DOOM game? The ability to play any DOOM file? But eventually it was decided that the purpose of Quandoom is not for anyone to actually play it and have fun, but instead to serve as a silly example of a classical algorithm applied in a quantum context.

\section{Classical Simulability}

There are a number of random checks in Quandoom: whether a shot hits its target, whether an enemy should be stunned by a hit, the actions of the enemies, among others. To do these, we apply a number of Hadamards onto a single register ``random'' at the very start of the circuit. This register is then compared to various values to flip bits with certain probabilities in order to do the aforementioned random checks.

It is known that Hadamard + Toffoli (H+T) is a universal quantum gate set \cite{aharonov2003simple}, meaning that any quantum circuit written in a different gate set can be converted to H+T using only a polynomial number of extra gates. As such, simulating H+T is theorized to be exponentially difficult in the worst case. Thus, how are we able to simulate Quandoom efficiently on a laptop?

The trick is that the random qubits are part of the ``garbage'' qubits, the last row of qubits in the screen qubits which are not used for rendering but take advantage of the measurement and resetting of the screen register. By resetting our random qubits, we ensure that in each run we only ever apply Hadamards onto a $\ket{0}$ state, thus at no point in the circuit do negative phases every appear. By never having negative phases, we will never have any cancellation, thus our final state will contain a equiprobable mixture of $2^\text{number of Hadamards}$ states. Therefore we can efficiently sample one of the possible outcomes by simply going though the circuit applying the gates to our initial state, then whenever we need to apply a Hadamard we just flip the qubit with $50\%$ probability. Based on this, we end with a conjecture:

\textbf{Conjecture:} A circuit consisting of only Hadamard and Toffoli gates is classically simulable in polynomial time if, and only if, it has an equivalent representation such that the Hadamards act on $\ket{1}$ states at most a logarithmic number of times.

\section{Acknowledgments}

Thanks to Raja Yehia and Marcio Taddei for being the first people to play it, and thanks to Ikko Hamamura for being the first person to look in detail at the QASM.

I don't put any funding sources because I only worked on this on weekends, this was a personal project and has nothing to do with my PhD or ICFO. I don't want people to think I wasted funding on this nonsense.

\bibliographystyle{unsrt}
\bibliography{refs}

\end{document}